\pdfoutput=1
\documentclass[twocolumn,aps,prl,reprint,amsmath,amssymb,floatfix]{revtex4-1}
\usepackage{graphicx}
\usepackage{bm}
\usepackage{color}

\begin{document}

\title{Origin of spin reorientation transitions in antiferromagnetic MnPt-based alloys}

\author{P.-H. Chang}

\author{I. A. Zhuravlev}

\author{K. D. Belashchenko}

\affiliation{Department of Physics and Astronomy and Nebraska Center for Materials and Nanoscience, University of Nebraska-Lincoln, Lincoln, NE 68588, USA}
\begin{abstract}
Antiferromagnetic MnPt exhibits a spin reorientation transition (SRT) as a function of temperature, and off-stoichiometric Mn-Pt alloys also display SRTs as a function of concentration. The magnetocrystalline anisotropy in these alloys is studied using first-principles calculations based on the coherent potential approximation and the disordered local moment method. The anisotropy is fairly small and sensitive to the variations in composition and temperature due to the cancellation of large contributions from different parts of the Brillouin zone. Concentration and temperature-driven SRTs are found in reasonable agreement with experimental data. Contributions from specific band-structure features are identified and used to explain the origin of the SRTs.
\end{abstract}

\maketitle

Antiferromagnetic materials are of interest for magnetoelectronic applications thanks to their insensitivity to stray fields and the accessibility of ultrafast dynamics. In particular, memory cells controlled by current-induced spin-orbit torque \cite{Jungwirth16}, terahertz spin-Hall nano-oscillators \cite{Cheng2016:PRL}, and magnetoelectric memory cells \cite{He2010,Kosub} have been proposed. These features also make antiferromagnets attractive for magnonic applications \cite{Daniels2015,Troncoso2015}.

Magnetocrystalline anisotropy energy (MAE) is an important parameter for antiferromagnonic devices, because it controls the spin wave spectrum at long wavelengths. Resonant parametric excitation by utilizing voltage-controlled MAE, like it was demonstrated for ferromagnets \cite{Verba2014,Chen2017}, could be used to generate spin wave packets in ultrathin antiferromagnetic nanostrips. Antiferromagnets with a small but tunable MAE are desirable to take advantage of the linear magnon spectrum for low-distortion signal transmission, while allowing for efficient spin wave generation, manipulation, and detection.

MnPt and off-stoichiometric alloys based on this tetragonal compound exhibit spin reorientation transitions (SRT) driven by both temperature and composition \cite{Kren1968JAP,Kren1968PR}, suggesting that they may be suitable for magnonic applications. While most measurements obtained easy-axis anisotropy \cite{Kren1968JAP,Kren1968PR,YorihikoJPS2010,Hiroaki_sflip} at room temperature, in-plane anisotropy has also been reported \cite{Stassis1979}. Although the magnetic moments on the Pt atoms vanish by symmetry, spin-orbit coupling on Pt can strongly influence MAE through hybridization with Mn, as in similar L1$_0$-ordered antiferromagnets \cite{Khmel2011}. Nevertheless, first-principles calculations find a small MAE ($K\sim 0.1$ meV/f.u.), which is more than an order of magnitude smaller compared to FePt, and different computational methods disagree in its sign \cite{Butler2010,Sakuma2006}. It was also found that MAE is very sensitive to band filling in the rigid-band approximation \cite{Sakuma2006}. All of these experimental and theoretical results clearly indicate a small and easily tunable MAE.

In itinerant magnets, anomalies in the temperature dependence of MAE may occur due to a variety of band-structure effects, such as the variation in band filling and band broadening induced by thermal spin fluctuations \cite{Zhuravlev2015}. Understanding of these effects calls for a first-principles analysis. In this paper, we examine the concentration and temperature dependence of MAE in MnPt-based alloys, obtaining the phase diagram in reasonable agreement with experimental data. Similar to the case of ferromagnetic (Fe$_{1-x}$Co$_x$)$_2$B alloys \cite{Zhuravlev2015}, we find that the temperature-induced SRTs observed in antiferromagnetic MnPt-based alloys are attributable to the effects of thermal spin disorder on the electronic structure.

Calculations were performed using the Green's function-based formulation of the tight-binding linear muffin-tin orbital (GF-LMTO) method and the coherent potential approximation (CPA) to describe substitutional disorder \cite{Turek2}. A series of Mn$_{1-x}$Pt$_{1+x}$ alloys was considered, where $x=0$ corresponds to the L1$_0$-ordered stoichiometric compound MnPt, while finite $x$ corresponds to excess Pt or Mn substituting randomly on the other sublattice. Concentration-dependent room-temperature lattice constants \cite{Kren1968PR} were smoothly interpolated and used in all calculations. We have verified that the results are not strongly affected by using temperature-dependent lattice constants for stoichiometric MnPt \cite{Kren1968PR}.

Thermal spin fluctuations were included on the same footing with substitutional disorder using the disordered local moment (DLM) method \cite{DLM1,DLM2,Staunton2004,Zhuravlev2015}. Integration over the orientations of Mn spins  was performed using a 122-point quadrature including 12 vertices, 20 face centers, and 30 edge centers of an icosahedron plus 60 vertices of a truncated icosahedron (buckyball). The quadrature weights were chosen so that any linear combination of angular harmonics with $l\leq15$ is integrated exactly. The statistical probability distribution for the spin orientations was taken from the mean-field approximation for the classical Heisenberg model at the given $T/T_N$ ratio, where $T_N$ is the N\'eel temperature.

Spin-orbit coupling (SOC) was included as a perturbation to the LMTO potential parameters \cite{Turek1,Belashchenko2015}, and the generalized gradient approximation \cite{GGA} was used for exchange and correlation. The anisotropy energy $K$ is calculated as the single-particle energy difference between the in-plane (100) and out-of-plane (001) orientations of the spins, taking the charge density from the self-consistent calculation without SOC. A uniform $32\times32\times32$ $k$-space mesh provided sufficient convergence for the Brillouin zone integration. The computational details are similar to Refs. \onlinecite{Zhuravlev2015,Belashchenko2015}. In the analysis of $\mathbf{k}$-resolved MAE, the data are symmetrized with respect to the $C_4$ rotation.

First, we study the influence of off-stoichiometry $x$ in Mn$_{1-x}$Pt$_{1+x}$ alloys on MAE at zero temperature. At $x<0$, the excess Mn atoms occupy the sites on the Pt sublattice, and their net interaction with the spins on the Mn sublattice vanishes by symmetry. Therefore, the spins of excess Mn atoms are expected to remain disordered if their concentration is small and the temperature is not very low. At larger concentrations, the interactions among the excess Mn spins could promote their ordering. To estimate its importance for MAE, we considered three hypothetical cases for excess Mn spins: fully ordered ferromagnetic, fully ordered antiferromagnetic, and fully disordered.

\begin{figure}[hbt]
\includegraphics[width=0.85\columnwidth]{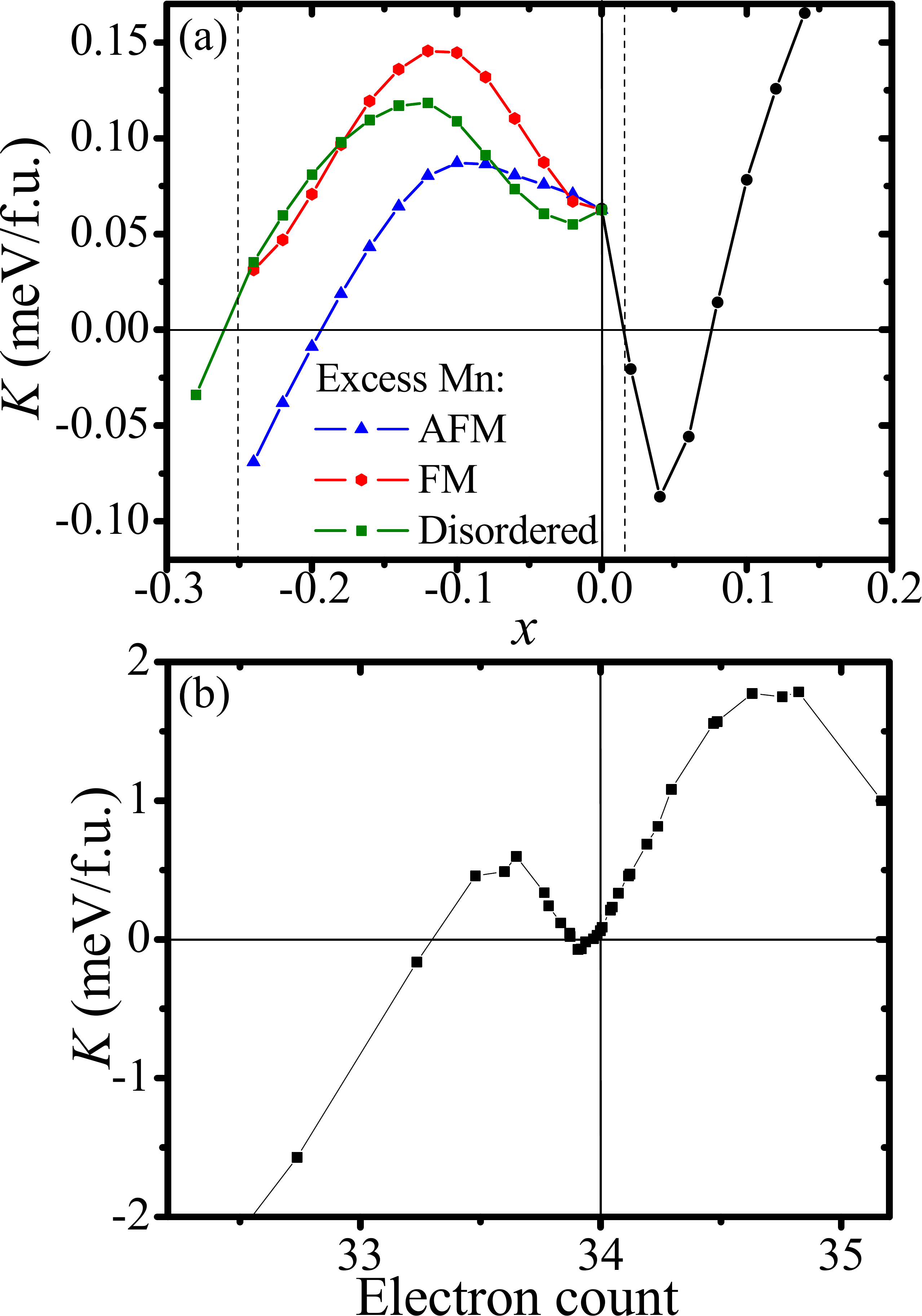}
\caption{\label{fig:mae_vs_x} (a) Calculated MAE of Mn$_{1-x}$Pt$_{1+x}$ at zero temperature. Three curves at $x<0$ correspond to different orderings of excess Mn spins (see text). The experimental SRTs \cite{Kren1968PR} are shown by vertical dashed lines. (b) Calculated MAE as a function of electron count per unit cell in the rigid-band model. The range of the electron count matches the range of $x$ in panel (a).}
\end{figure}

Fig. \ref{fig:mae_vs_x}(a) shows the MAE calculated in CPA as a function of concentration. We see that MAE exhibits a qualitatively similar behavior for all three descriptions of excess Mn. Therefore, we did not attempt to determine the ground state but simply considered the disordered configuration in all calculations that follow.

In agreement with experimental data \cite{Kren1968PR,Kren1968JAP}, Fig.\ \ref{fig:mae_vs_x}(a) shows easy-axis anisotropy at $T=0$ in stoichiometric MnPt and SRTs to easy-plane anisotropy at $x\approx -0.26$ and $x\approx 0.02$ \cite{note-sign}. In addition, we find another SRT back to easy-plane anisotropy at $x\approx0.08$, which, to our knowledge, has not been experimentally reported.

Fig. \ref{fig:mae_vs_x}(b) shows the MAE as a function of the electron count in the rigid-band approximation, which agrees well with the calculations of Refs.\ \onlinecite{Sakuma2006}. However, the comparison with CPA calculations in Fig.\ \ref{fig:mae_vs_x}(a) clearly shows that the rigid-band approximation fails to describe the behavior of MAE in Mn$_{1-x}$Pt$_{1+x}$ alloys even on the qualitative level. In particular, it predicts a qualitatively wrong behavior of MAE on the Pt-rich side and overestimates it by an order of magnitude. Full CPA calculations are, therefore, essential for the description of MAE in this system.

If SOC is treated as a perturbation to the Hamiltonian, the second-order approximation for MAE can be represented as a sum of pairwise contributions corresponding to the pairs of sites on which the two SOC operators are applied in the perturbative expansion \cite{Solovyev}. This description is approximate, because second-order perturbation theory can fail in metals for band crossings near the Fermi level, especially in antiferromagnets where all bands are degenerate by spin. In FePt and CoPt the MAE is dominated by single-site terms on Pt \cite{Solovyev}. To estimate the role of different terms for antiferromagnetic MnPt, we performed two auxiliary calculations with SOC suppressed on Mn or Pt atoms. The MAE is negligibly small if SOC on Pt is suppressed, but it is large and positive (1.1 meV/f.u.) if SOC is suppressed on Mn. This indicates that, in contrast to FePt and CoPt, in MnPt the large negative Mn-Pt cross-term nearly cancels the large positive term coming solely from SOC on Pt.

\begin{figure}[htb]
\centerline{\includegraphics[width=0.95\columnwidth]{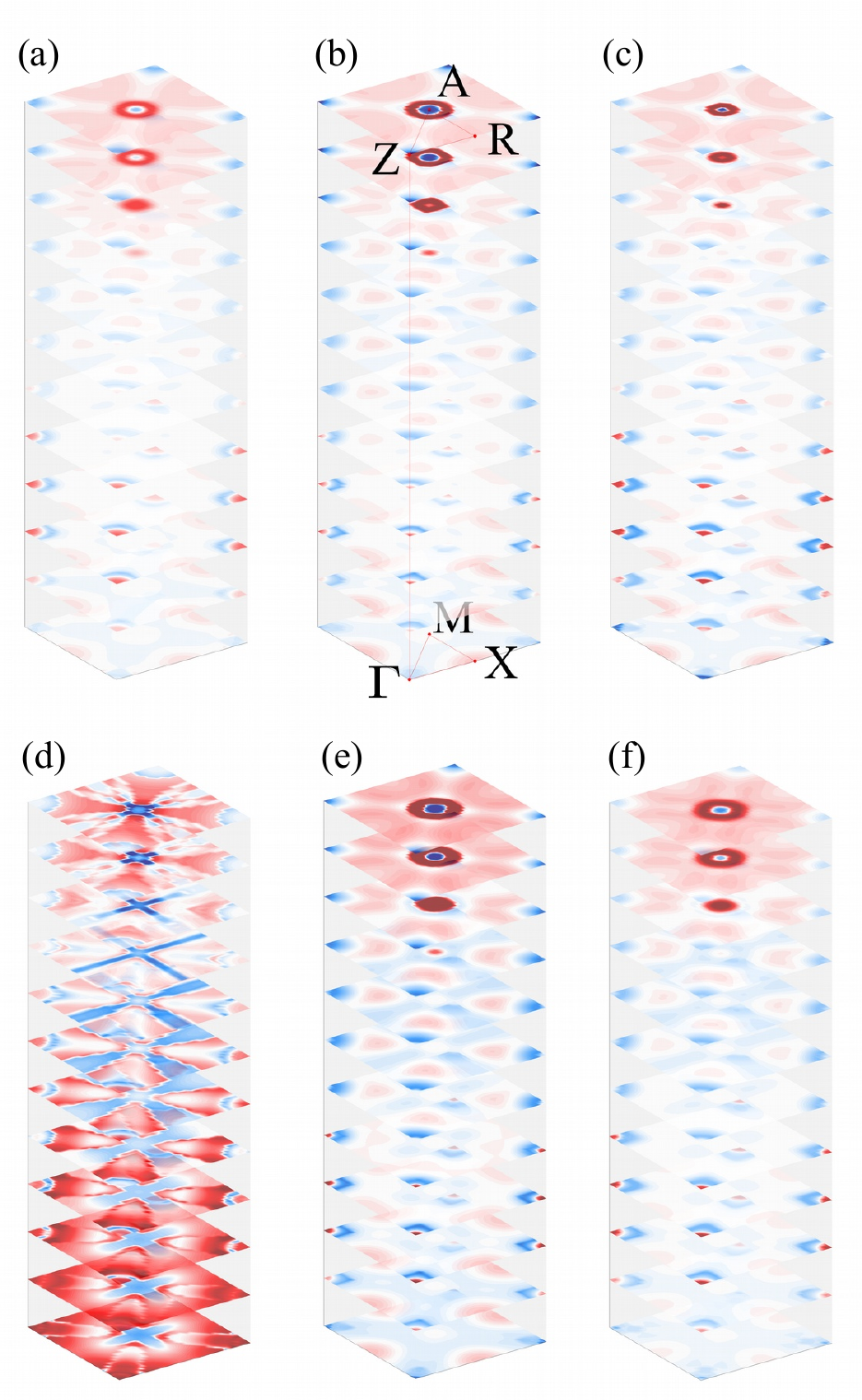}}
\caption{\label{fig:kmcax} $C_4$-symmetrized $\mathbf{k}$-resolved MAE in Mn$_{1-x}$Pt$_{1+x}$ at (a) $x=-0.12$, (b) $x=0$, (c) $x=0.04$, (d) in stoichiometric FePt, (e) in pure MnPt at $T/T_N=0.25$ (near the SRT) and (f) $T/T_N=0.75$ (near the minimum of MAE). Red (blue) color shows positive (negative) values; the range of values in panels (e)-(f) is 2.5 narrower compared to panels (a)-(d).}
\end{figure}

To obtain insight into the origin of the SRTs, let us examine the $\mathbf{k}$-resolved MAE displayed in Fig.\ \ref{fig:kmcax}. Panel (b) for stoichiometric MnPt shows large positive and negative contributions from different regions of the Brillouin zone, which largely cancel each other out. For comparison, panel (d) for ferromagnetic FePt shows that positive contributions dominate over most of the Brillouin zone in that material, adding up to a large easy-axis anisotropy of about 2.6 meV/f.u., in agreement with other calculations \cite{Butler2010,Sakuma1994,Shick2003}. Note that a very large easy-axis anisotropy of 4.7 meV/f.u.\ was obtained for the hypothetical ferromagnetic phase of MnPt \cite{Ravindran} at the experimental lattice parameters. Strong dependence of MAE on the magnetic state was also found in other similar compounds \cite{Khmel2011}.

\begin{figure}
\centerline{\includegraphics[width=0.95\columnwidth]{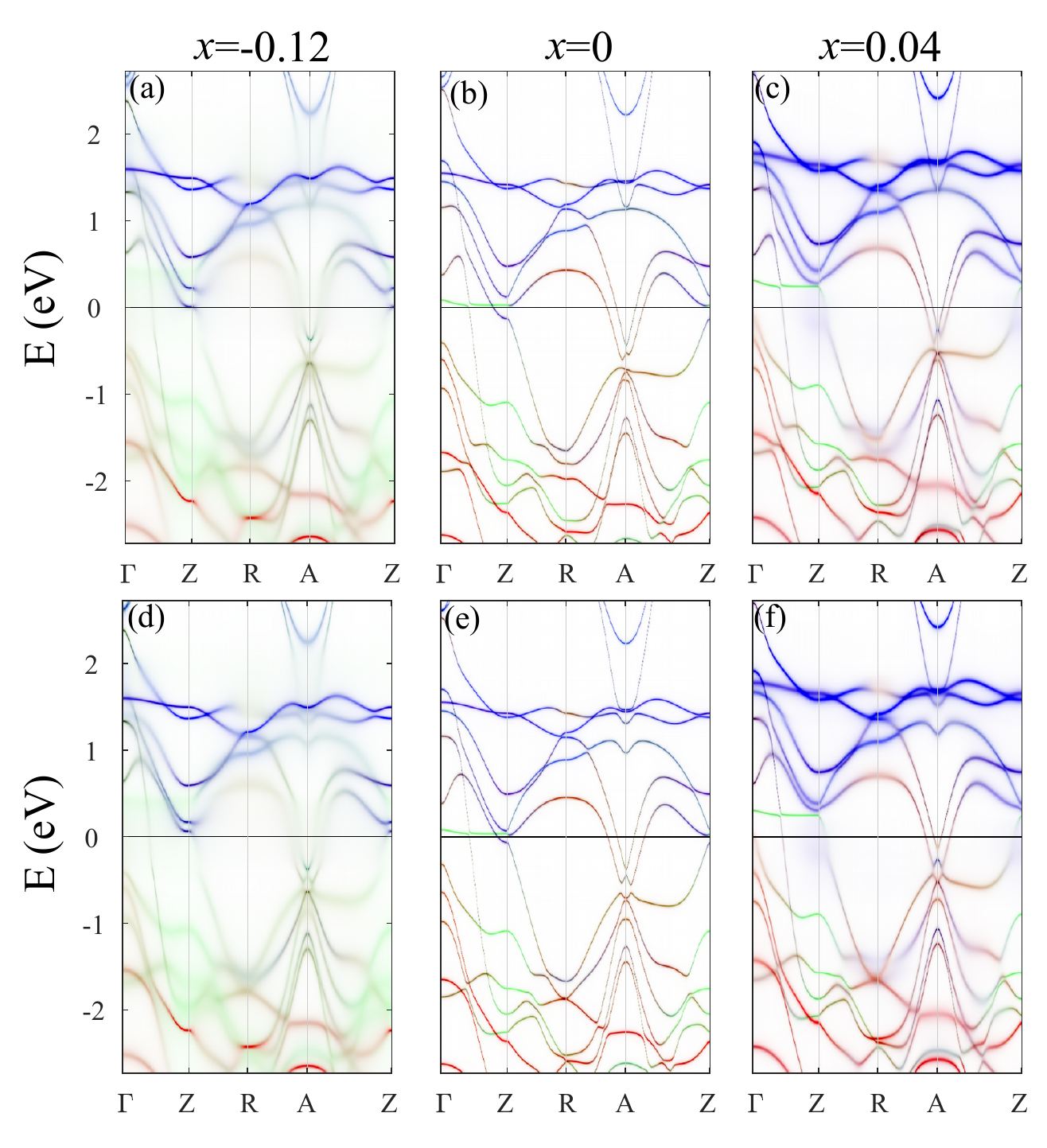}}
\caption{\label{fig:spect_T0} Bloch spectral functions in Mn$_{1-x}$Pt$_{1+x}$. Red and blue color densities represent the spectral weights of the majority and minority-spin states of Mn, and green represents all states of Pt. The order parameter is oriented along the (100) direction in panels (a)-(c) and along (001) in panels (d)-(f). The concentrations $x$ are indicated above the panels.
}
\end{figure}

One can identify three distinct features in Fig.\ \ref{fig:kmcax}(a)-(c): the strong spherical ``hot spot'' giving mostly positive contribution around the $A$ point, the cylindrical region around the $\Gamma Z$ line giving a negative contribution, and the slowly varying background. The comparison of panels (a), (b), and (c), corresponding to different concentrations, suggests that the $A$ hot spot and the $\Gamma Z$ cylinder are sensitive to off-stoichiometry.

Sharp features in $\mathbf{k}$-resolved MAE come from the pairs of occupied and unoccupied bands near the Fermi level that are strongly mixed by SOC \cite{Kondorskii,Solovyev}.
Fig. \ref{fig:spect_T0} shows Bloch spectral functions, for the same concentrations as in Fig.\ \ref{fig:kmcax}(a)-(c), at $T=0$, along several high-symmetry lines in reciprocal space, for two orientations of the antiferromagnetic order parameter: (100) for panels (a)-(c) and (001) for (d)-(f). Centered around the $A$ point, we find two bands with conical dispersions crossing the Fermi level. These two bands are degenerate in the absence of SOC, and their splitting depends on the orientation of the order parameter, producing the hot spot in $\mathbf{k}$-resolved MAE around the $A$ point in Fig.\ \ref{fig:kmcax}.

At the $Z$ point, SOC also splits two otherwise degenerate bands, pushing one of them above and the other below the Fermi level. This results in a hot spot at $Z$, which makes a negative contribution to MAE. The mixing of other bands along the $\Gamma Z$ line also results in a pronounced negative contribution to MAE. The sensitivity of the sharp features seen in Fig.\ \ref{fig:kmcax} to off-stoichiometry can be traced to the changing occupations, hybridization, and broadening of the bands in Fig. \ref{fig:spect_T0}.

The smooth background in Fig.\ \ref{fig:kmcax}(a)-(c) comes from the mixing of the occupied and unoccupied states that are far away from the Fermi energy.

It is interesting to note that the bands dominated by Pt states (seen as green in Fig.\ \ref{fig:spect_T0}) are insensitive to the orientation of the order parameter, which is because Pt atoms carry no magnetic moments. However, spin-orbit coupling on Pt contributes to MAE through its effect on the hybridized bands carrying both Mn and Pt character. This is a reciprocal-space counterpart of the argument of Ref.\ \cite{Khmel2011} for similar L1$_0$-ordered MnX antiferromagnets, which interpreted the contribution of SOC on X atoms to MAE in terms of the non-trivial real-space spin-density distribution on X. In contrast, the Pt atoms are magnetically polarized in \emph{ferro}magnetic compounds like FePt and CoPt, whereby all bands can contribute to MAE.

\begin{figure}[htb]
\centerline{\includegraphics[width=0.9\columnwidth]{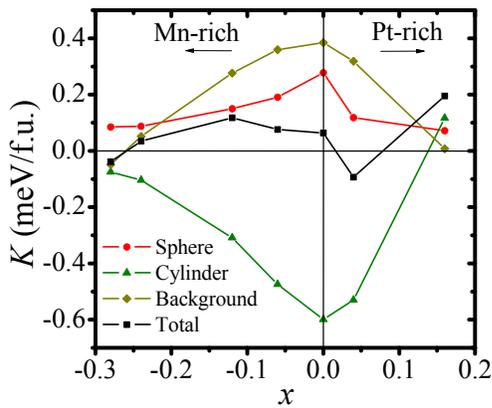}}
\caption{\label{fig:decomp_mca_x} Contributions to MAE from different parts of the Brillouin zone as a function of off-stoichiometry.}
\end{figure}

Fig.\ \ref{fig:decomp_mca_x} shows the concentration dependence of the contributions to MAE integrated over three regions discussed above: the sphere enclosing the hot spot near $A$, the cylinder containing the features near the $\Gamma Z$ line, and the rest of the Brillouin zone (background). All three contributions are fairly large and of the same order of magnitude, but there is a strong cancellation of positive and negative contributions. All contributions are reduced by off-stiochiometry (both excess Mn and excess Pt) but at different rates, leading to large relative variations in the total MAE and to the SRTs. The strong cancellation also makes MAE at small $x$ sensitive to the variations in the $\mathit{c/a}$ ratio \cite{Sakuma2006,sakuma2006-1,Butler2010} and to temperature changes, as we show below.

\begin{figure}[htb]
\centerline{\includegraphics[width=0.85\columnwidth]{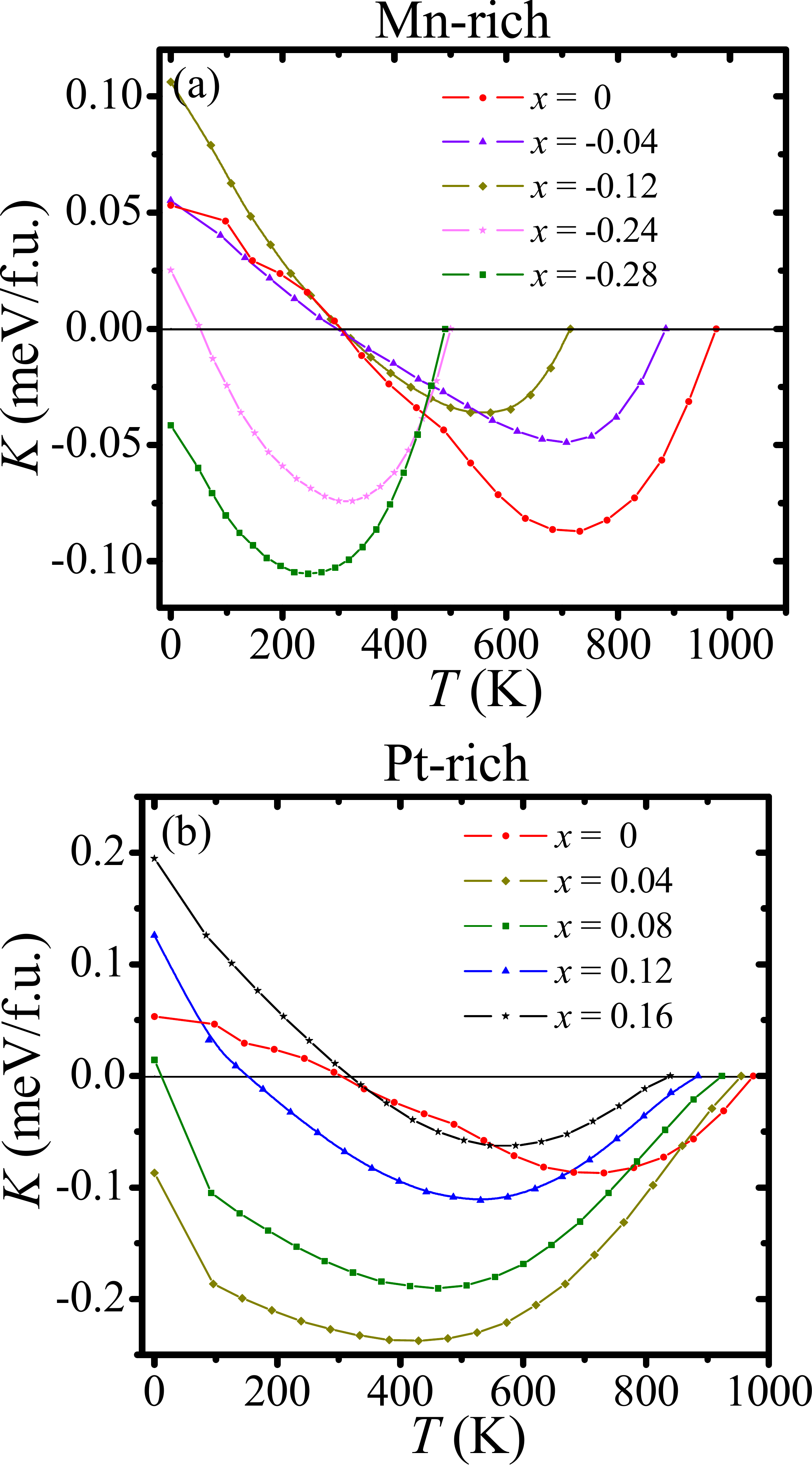}}
\caption{\label{fig:mca_t} Temperature dependence of MAE for (a) Mn-rich and (b) Pt-rich Mn-Pt alloys.}
\end{figure}

The temperature dependence of MAE, calculated for different concentrations using the CPA-DLM method, is displayed in Fig. \ref{fig:mca_t}, where we used the experimental N\'eel temperature \cite{Kren1968PR} for each concentration. The temperature dependence of MAE is anomalous (non-monotonic) at all concentrations: the MAE decreases at low temperatures but then passes through a minimum and increases back to zero as the temperature tends to the N\'eel point. Thus, at those concentrations where $K$ is positive (easy-axis) at $T=0$, we always find a SRT.

\begin{figure}[htb]
\centerline{\includegraphics[width=0.9\columnwidth]{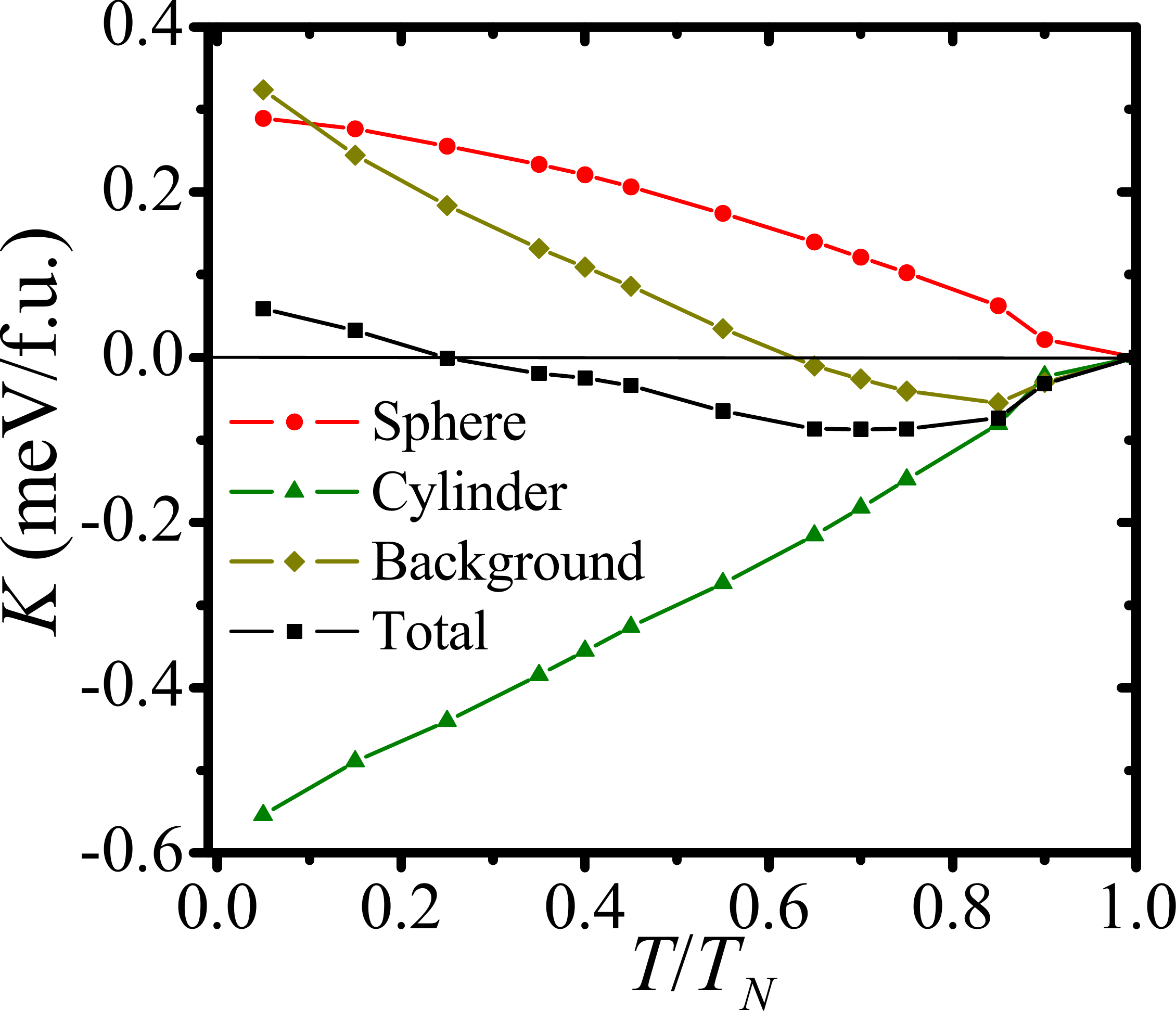}}
\caption{\label{fig:decomp_mca_t} Temperature dependence of the contributions from different parts of the Brillouin zone in stoichiometric MnPt.}
\end{figure}

\begin{figure}[htb]
\centerline{\includegraphics[width=0.95\columnwidth]{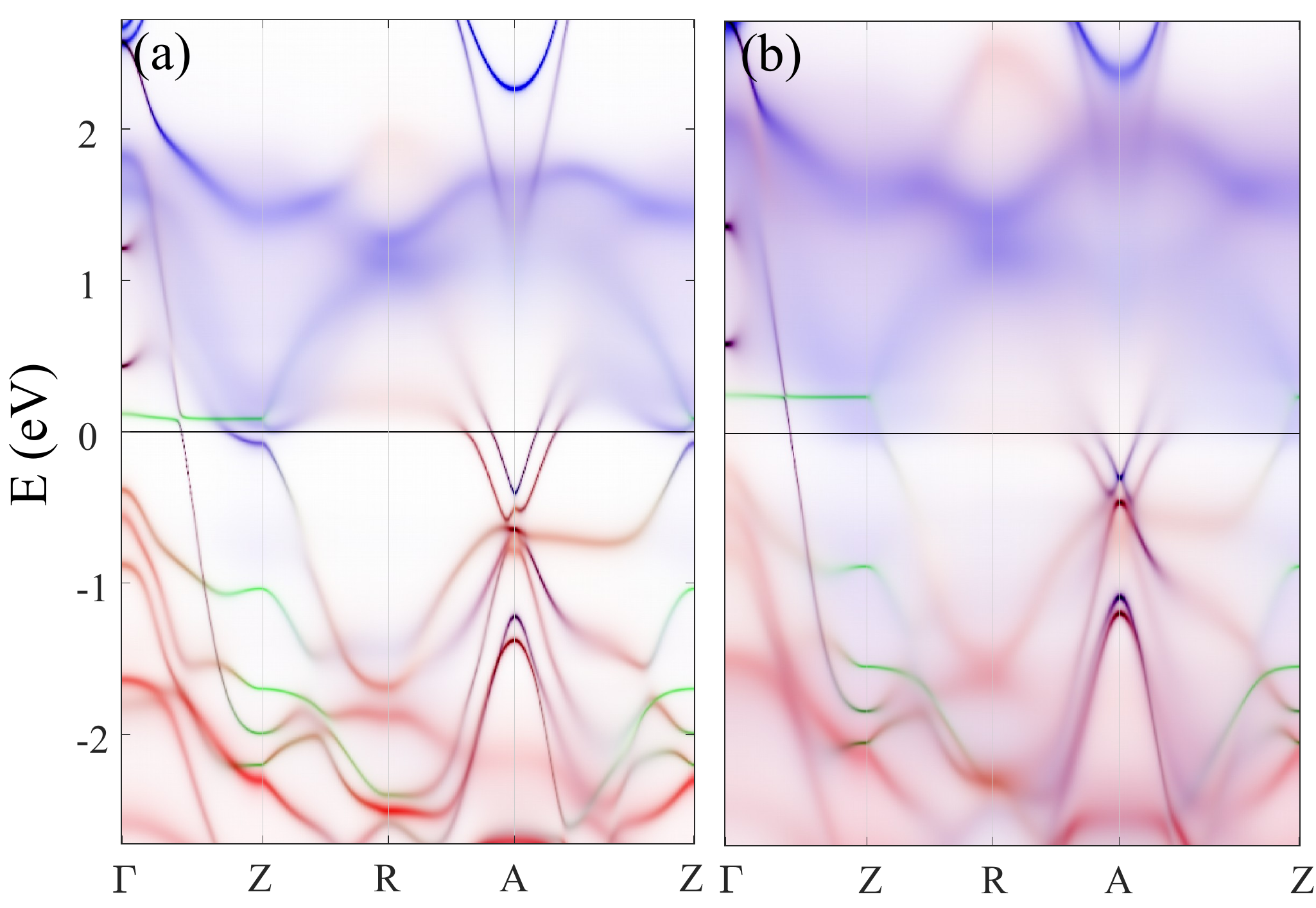}}
\caption{\label{fig:spect_dlm} Spectral functions of stoichiometric MnPt at (a) $T/T_N=0.25$ and (b) $T/T_N=0.75$. }
\end{figure}

As we saw in Figs.\ \ref{fig:kmcax}-\ref{fig:decomp_mca_x} above, the electronic bands crossing or approaching the Fermi level in different regions of the Brillouin zone produce large contributions of opposite signs to MAE, which are sensitive to the shifts and broadening of those bands. For stoichiometric MnPt, the temperature dependences of the $\mathbf{k}$-resolved MAE, integrated contributions from different parts of the Brillouin zone, and spectral functions are displayed in Figs. \ref{fig:kmcax}(e)-(f), \ref{fig:decomp_mca_t}, and \ref{fig:spect_dlm}, respectively.

Figs. \ref{fig:kmcax}(b),(e)-(f) and \ref{fig:decomp_mca_t} show that the positive background contribution [smooth red regions in Fig.\ \ref{fig:kmcax}(b),(e)-(f)] decreases the fastest with increasing temperature, to the extent that the entire background contribution turns negative at $T/T_N\approx0.6$. Note that the color map in panels (e)-(f) has been rescaled to emphasize this background. In contrast, the contribution from the vicinity of the $A$ point decreases relatively slowly with temperature. The competition of large contributions declining at different rates results in the anomalous behavior of the total MAE and leads to a SRT.

Fig.\ \ref{fig:spect_dlm} shows that some of the bands, including those that are split by SOC near $Z$, are strongly broadened already at $T/T_N=0.25$, and most bands are completely smeared out at $T/T_N=0.75$. On the other hand, the conical bands around the $A$ point are visible even at $T/T_N=0.75$, which explains the slow decline of their contribution to MAE.

\begin{figure}[htb]
\centerline{\includegraphics[width=0.9\columnwidth]{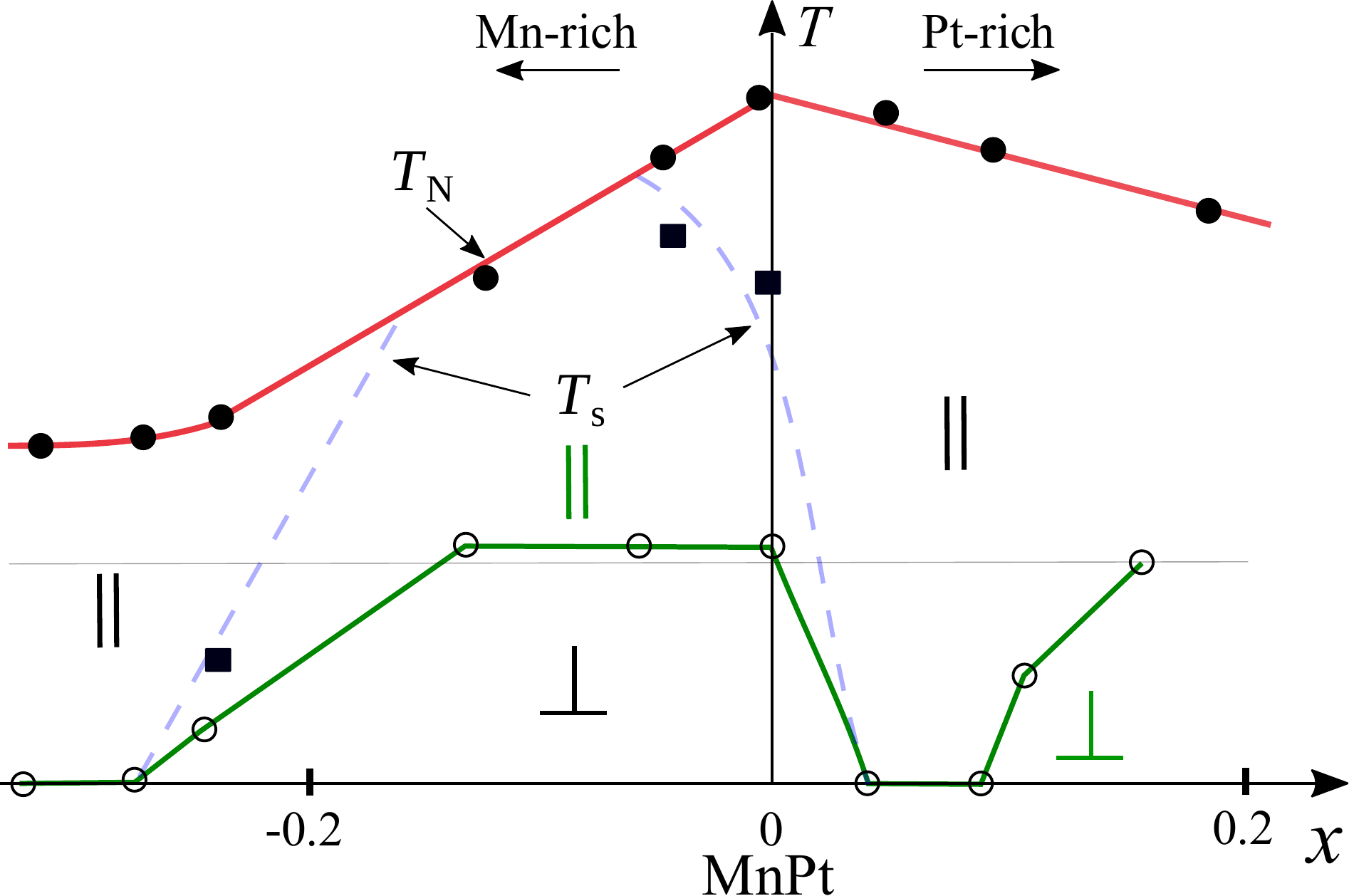}}
\caption{\label{fig:phase} Predicted phase diagram using the calculated $T_s/T_N$ ratios and the experimental data for $T_N$ (black filled circles guided by the solid red line) \cite{Kren1968PR}. Black filled squares: experimental SRT $T_s$ \cite{Kren1968PR}; dashed blue line: sketch of the experimental $T_s(x)$. Open circles connected by solid green line: theoretical $T_s$.}
\end{figure}

The phase diagram based on our results is plotted in Fig. \ref{fig:phase}, which also shows the experimental data \cite{Kren1968PR}. Note that we did not attempt to determine the N\'eel temperature $T_N$, because our focus is on understanding the behavior of the MAE. The CPA-DLM calculations (Fig.\ \ref{fig:mca_t}) produce $T_s/T_N$, where $T_s$ is the temperature of the SRT.

As discussed above in connection with Fig.\ \ref{fig:mae_vs_x}(a), the SRTs predicted at $x\approx -0.26$ and $x\approx 0.02$ at zero temperature agree with experimental results. Our calculation predicts a thermal SRT in the entire range $-0.26<x<0.02$. In experiment \cite{Kren1968PR,note-sign}; a thermal SRT was found at $x=0$, $x=-0.04$, and $x=-0.24$, but not at $x=-0.13$. In addition, the SRTs predicted by CPA-DLM occur at considerably lower temperatures compared to experiment. These quantitative differences are not surprising in view of the strong cancellations of different contributions to MAE. In particular, an underestimated MAE at $T=0$ would also lead to an underestimated SRT temperature $T_s$. In addition, we note that the observed SRT (detected using the [101]/[100] Bragg peak intensity ratio) occurs in a fairly wide temperature range \cite{Kren1968PR,Hiroaki_sflip}. A careful study using a stoichiometric MnPt single crystal \cite{Hiroaki_sflip} reported a gradual transition between 580 and 770 K. Based on the analysis of the magnon spectrum, it was concluded that the transition involves the changing volume ratios of the easy-axis and easy-plane regions. This finding suggests that the gradual character of the SRT is associated with the spatial inhomogeneity of the L1$_0$ order parameter or concentration. Such variations are likely to be even larger in the off-stoichiometric powder samples \cite{Kren1968PR}. On the other hand, our CPA-DLM calculations assume a perfectly homogeneous alloy with the maximal order parameter allowed at the given concentration.

To conclude, we have studied the concentration and temperature dependence of magnetocrystalline anisotropy in L1$_0$-ordered Mn-Pt alloys using first-principles calculations. The strong cancellation of contributions from different regions in the Brillouin zone explains the small magnitude of the anisotropy energy and its sensitivity to off-stoichiometry and temperature changes, which gives rise to concentration and temperature-driven spin reorientation transitions.

We are grateful to Sergii Khmelevskyi and Ilya Krivorotov for useful discussions. This work was supported by the Nanoelectronics Research Corporation (NERC), a wholly-owned subsidiary of the Semiconductor Research Corporation (SRC), through the Center for Nanoferroic Devices (CNFD), a SRC-NRI Nanoelectronics Research Initiative Center (Task ID 2398.003), and by the National Science Foundation through the Nebraska MRSEC (Grant No. DMR-1420645) and Grant No. DMR-1609776. Calculations were performed utilizing the Holland Computing Center of the University of Nebraska, which receives support from the Nebraska Research Initiative.

\end{document}